# Fundamental Scaling Relationships in Additive Manufacturing and their Implications for Future Manufacturing Systems

David M. Wirth[1*], Chi Chung Li[2], Jonathan K. Pokorski[1], Hayden K. Taylor[2], and Maxim Shusteff[3]*

[1]Department of NanoEngineering, University of California San Diego, Jacobs School of Engineering, La Jolla, California 92093, United States
[2]Department of Mechanical Engineering, University of California, Berkeley, Berkeley, CA 94720, United States
[3]Lawrence Livermore National Laboratory, 7000 East Avenue, Livermore, California 94550, United States
*Corresponding Author(s)
Email Address: dwirth9@gmail.com
Phone Number: 424-256-5087
Address: 9500 Gilman Dr., SME Building 243J, La Jolla, California 92093, United States
Email Address: shusteff1@llnl.gov
Phone Number: 925-423-0733
Address: 7000 East Avenue, Livermore, California 94550, United States

Abstract:

The field of additive manufacturing (AM) has advanced considerably over recent decades through the development of novel methods, materials, and systems. However, as the field approaches maturity, it is relevant to investigate the scaling frontiers and fundamental limits of AM in a generalized sense. Here we propose a simplified universal mathematical model that describes the essential process dynamics of many AM hardware platforms. We specifically examine the influence of several key parameters on total manufacturing time, comparing these with performance results obtained from real-world AM systems. We find a inverse-cubic dependency on minimal feature size and a linear dependency on overall structure size. These relationships imply how certain process features such as parallelization and process dimensionality can help move toward the fundamental limits. AM methods that are capable of varying the size of deposited voxels provide one possibility to overcome these limits in the future development of AM. We also propose a new framework for classifying manufacturing processes as "top-down" vs "bottom-up" paradigms, which differs from the conventional usage of such terms, and present considerations for how "bottom-up" manufacturing approaches may surpass the fundamental limits of "top-down" systems.

## I. Table of Symbols

| Symbol | | Description | Units |
|---|---|---|---|
| $T$ | - | Time required to print an object | [s] |
| $s$ | - | Average toolpath velocity | [m/s] |
| $r_x$ | - | Toolpath width / minimum feature size / voxel length | [m] |
| $r_t$ | - | Layer thickness | [m] |
| $d$ | - | Toolpath aspect ratio ($r_t/r_x$) | [*] |
| $v_0$ | - | Voxel volume | [m$^3$] |
| $V_T$ | - | Total object volume | [m$^3$] |
| $Q$ | - | Volumetric build rate | [m$^3$/s] |
| $t_0$ | - | Average time required for the toolpath to move a distance $r_x$ | [s] |
| $c$ | - | Scaling constant | [*] |
| $n$ | - | Number of simultaneous toolpaths | [*] |
| $k$ | - | "Printer constant" | [*] |
| VPR | - | Voxel Patterning Rate | [Hz] |
| $t_d$ | - | Doubling time of a self-replicating system | [s] |
| $m$ | - | Geometric constant of a self-replicating system | [*] |

\* Dimensionless

## II. Background/Introduction:

Additive manufacturing (AM), also known as 3D printing, is most commonly defined as a process by which material is added to an object, specified by digital data from a 3D model, typically in a layer-by-layer fashion. AM is usually contrasted with subtractive manufacturing methods, in which material is removed (cut, abraded, eroded, vaporized, etc) from the bulk of a larger volume. AM also differs from formative methods, in which one material serves as a mold to template the formation of another material. The mold is generally a copy or an inverse copy of the desired finished structure. Since Charles Hull's commercialization of the process known today as stereolithography (SLA) in 1984[1] , nearly four decades of progress have brought about a great variety of AM methods, printer systems, and materials.[2–4] Given the degree to which these technologies have matured, it has become possible to identify common principles and extract generalized parameters that lend themselves towards a universal description of such systems.

We are interested here in taking stock of the full range of AM approaches to elucidate the fundamental limitations that may constrain their further development. We therefore propose a universal framework for describing the speed of AM as it scales with feature resolution and overall structure size. This allows the evaluation of the inherent trade-offs and limitations that are implied by this universal description, and its consequences for the scalability of AM as a manufacturing paradigm, and ultimately for the field of manufacturing as a whole. We find that those AM approaches which are not fully described by our model provide clues to possible directions for future research and development to overcome the inherent limitations of the technique.

In the most general sense, manufacturing is a process by which information is converted into ordered matter. This process involves three interrelated sub-processes, which are 1. memory retrieval/information processing, 2. hardware operation, and 3. material response. For each sub-process, a maximum rate or throughput can be determined, and the slowest of them provides the limiting function (or "bottleneck") for the speed of the overall manufacturing process. We can informally define these sub-processes as follows: Memory retrieval refers to accessing stored structural information and information processing refers to (if necessary) decompressing or otherwise converting that information to a format that is usable by the hardware. Hardware operation rate is how fast the deposition machine or assembler can respond to commands by the information retrieval/processing system and respond to those commands to effect material change (addition, removal, or other substantial change). The material response describes the speed at which the material can accept this information and respond to it, undergoing deformation, flow, melting, solidification or other required physico-chemical transition. The sub-processes of course influence each other, and each could be the topic of its own detailed review. Our purpose in this work is to investigate a generalized treatment of the second sub-process: hardware operation (leaving memory retrieval and material response for future work).

Specifically, we will first explore the rate of hardware operation in AM systems, as it is often the rate-limiting sub-process, and it has been studied and documented extensively across a wide range of AM approaches. We consider in particular how a maximum rate might be measured and key principles that drive scaling relationships for achieving larger and more complex structures.

### A. The Model Framework

To model a generalized AM apparatus, let us begin by considering an extruder nozzle (**Figure 1A/B**), which moves with a linear speed $s$ depositing material with a characteristic width $r_x$.

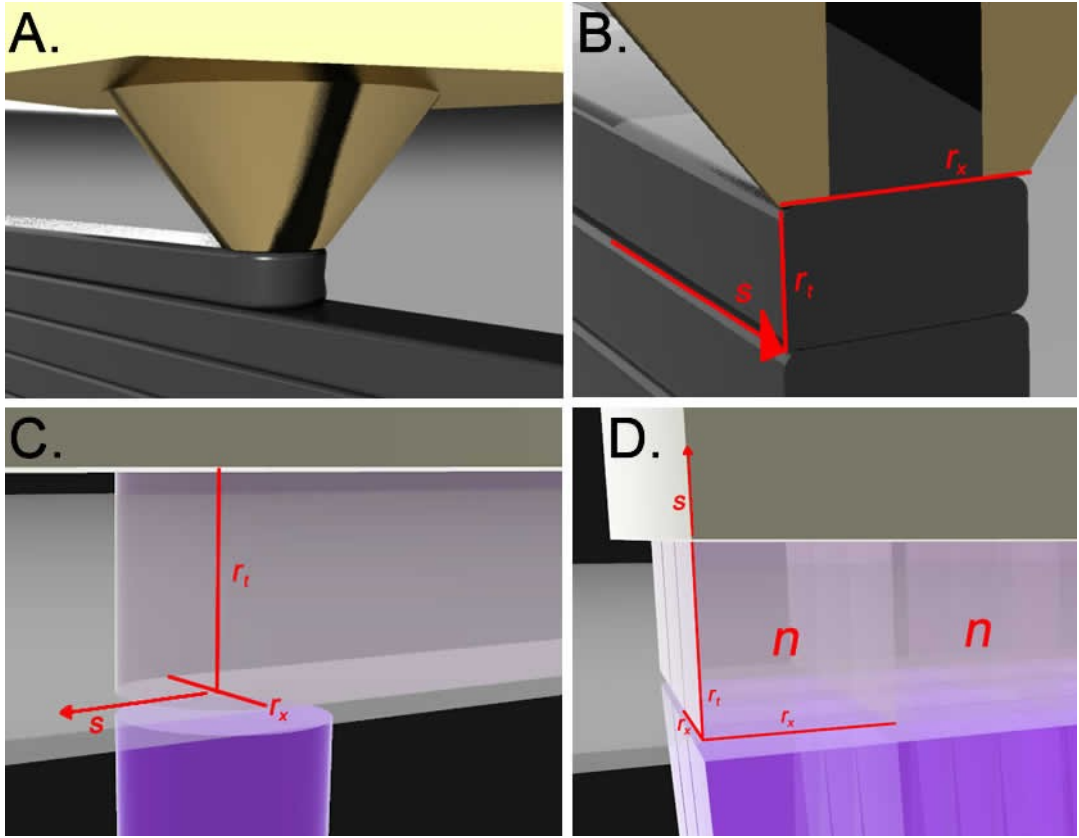

**Figure 1:** Notional diagrams of various AM processes including: A-B) FDM, C) SLA, and D) DLP/CLIP/MSLA and associated key geometric parameters noting the similarities between each process. Each process has ($n$) deposition paths, each depositing material in a volume of ($cr_t r_x^2$) at an average toolpath speed ($s$).

In layer-by-layer AM processing, the layer thickness $r_t$ is usually of a similar spatial size scale as $r_x$, and we will denote the aspect ratio of each parallelepipedal volume element (voxel) as $d = r_t/r_x$. Therefore, each voxel's volume is $v_0 = r_x^2 r_t = d r_x^3$, and the time needed for the toolpath to move its own width (i.e. the time to deposit each voxel) is $t_0 = r_x/s$. We may then express the volumetric build rate as $Q = d r_x^3 / t_0$ and the total time required to print an arbitrary object with a total volume $V_T$ can be written as:

$$T = \frac{V_T}{Q} = \frac{V_T t_0}{d r_x^3} \tag{1}$$

Many AM structures are designed with void space in them, taking advantage of lightweighting approaches[2,5], metamaterial properties[6–8], or other considerations. We therefore introduce a scaling constant $c$ that ranges from 0 to 1 to represent the average solid fraction, where $c = 1$ is the fully solid condition, and $c < 1$ for all structures with fractional solid fill (i.e. an infill percentage less than 100%). The volumetric build rate then becomes $Q = dr_x^3/ct_0$, since the total build volume per unit time can be larger when the toolhead does not need to visit every single voxel within the 3D volume. We note that this fill-factor scaling is most relevant for vector-based processes, typically those with a single toolpath, such as SLA and fused filament fabrication (FFF). In processes which are moderately parallelized, such as Polyjet/binder jetting[9,10], void space has a non-negligible effect on print time because some portions of the build space can still be "skipped". In highly parallelized processes such as digital light projection (DLP) and tomographic volumetric additive manufacturing (tomographic VAM), void space has no effect on print time, because the build process is unaffected by non-solid regions.

Since this description is derived from an extruder nozzle, it is valid for any extrusion-based methods such as FFF[11], direct ink writing (DIW)[12,13] and also bioprinting approaches that rely upon extrusion.[14] The description also holds for other scanned-toolpath AM methods, for which a representative example is laser-scanning based vat photopolymerization, such as SLA (**Figure 1C**). Here, although the fundamental physics of how material is being added to the printed structure are different than extrusion, our model still captures the essential behavior of a single toolpath creating a predefined pattern in space point-wise or line-by-line. The resin vat-based SLA category introduces another feature common across many AM processes, which is the time required to reflow the resin material before depositing the next layer. In selective laser sintering (SLS) or laser powder bed fusion (LPBF), this is the time needed to recoat a new layer of powder feedstock. Our model takes this recoating time into account by specifying that the toolpath speed $s$ is not the maximum instantaneous speed, but rather the effective speed of the toolpath *throughout the duration of the print.* The time spent travelling, recoating or moving between layers is taken into account when computing the effective speed for a given machine and exhibits slight variation depending on object geometry.

In some AM systems, multiple "toolpaths" are active simultaneously. The clearest example of this parallelization is projection-based vat photopolymerization, DLP or continuous liquid interface production (CLIP)[15] (**Figure 1D**), in which many projected image pixels are cured simultaneously, and Polyjet in which many nozzles deposit material in parallel. Newer methods which use multiple parallel toolpaths include Diode-based Additive Manufacturing (DiAM)[16] and multi-focal two-photon polymerization.[17,18] To model the behavior of such parallelized systems, we introduce one more scaling constant $n$ to represent the multiple deposition steps that occur simultaneously. Incorporating all of the scaling constants, the volumetric deposition rate then becomes $Q = ndr_x^3/ct_0$, and the total print time is:

$$T = \frac{V_T}{Q} = \frac{cV_T t_0}{ndr_x^3} \tag{2}$$

We can conclude our generalized treatment of AM processes by defining a "printer constant" as the quantity $k = ct_0/d$, which groups together nozzle geometric parameters and the speed of toolpath movement, and accounts for partial fill fraction and interlayer time. The printer constant is useful for grouping the individual parameters of a specific AM system and part geometry, representing the characteristic effective timescale to fabricate a single voxel. This allows us to arrive at a simplified version of **Eq. 2**, relating the time $T$ to print an object of volume $V_T$ using a deposition path with characteristic minimum dimension $r_x$:

$$T = \frac{kV_T}{nr_x^3} \tag{3}$$

This generalized result is perhaps unsurprising. We find that the total build time of an AM object scales linearly with its total volume, and scales inversely with the cube of the minimum printed feature size. This represents the best-case performance limit of AM systems, when other practical limitations are minimized or optimized. The presence of the scaling constants highlights the fact that an AM system can be made much faster by parallelization (large $n$, such as a DLP process exposing ~$10^6$ voxels simultaneously), or by making the single-voxel timescale very rapid (small $k$) with methods such as two photon polymerization (2PP) for which $t_0$ ~$10^{-6}$ s.

### B. Universality of our Scaling Relationship

We can evaluate the generality of this simple model beginning with the seven category groupings for AM processes defined by the International Standards Organization (ISO): 1. Binder jetting (BJT), 2. Directed energy deposition (DED), 3. Material extrusion (MEX), 4. Material jetting (MJT), 5. Powder bed fusion (PBF), 6. Sheet lamination (SHL), and 7. Vat photopolymerization (VPP).[19] Categories 1-5 are clearly well-represented by our description of a moving toolpath that scans through voxels of a characteristic size over the desired build volume. MEX corresponds to our initial description of an extrusion nozzle, while MJT is a discretized (and parallelized) version of this process depositing material drop-wise. Both BJT and PBF (categories 2 and 5) process powdered feedstock materials by sintering/melting or binding them together, where the laser focus or the binder nozzle define the working voxel. DED continually delivers new material by either aerosolized powder or wire feed[20], while a laser or electron beam fuses the incoming material to the workpiece. Thus DED manufacturing also uses a toolhead scanning through space depositing voxels of a characteristic size. SHL is the only one of the seven ISO categories that is not well described by our generalized model, and we will take up the implications of this in Section D below. Finally, the VPP process category requires somewhat more detailed examination, since it includes a highly diverse and evolving family of processes. It is apparent that our model holds for the various layer-by-layer methods we have already mentioned, but this is less clear for emerging volumetric approaches, which we will examine in the next section.

### C. Model Validation Against Real AM System Performance

To relate this generalized model to actual AM system performance, in **Figure 2A**, we plot the volumetric build rate of various classes of AM systems against the minimum feature size (i.e. spatial resolution) of such systems (the complete list of systems and source data is given in **Table 1**). This type of plot was first compiled by Shusteff *et al.*[21] to demonstrate the advantages of the volumetric printing approach. The overall shape of the trend is approximately cubic (i.e. slope of ~3 on a log-log plot), as suggested by **Eq. 3**. However, within each feature resolution class, the build rates span several orders of magnitude, which is due to the specifics of the deposition physics affecting the single-voxel timescale *k* as well as the scaling factors we discussed above. While efforts have been made to compile a representative swath of data from disparate AM systems, the data contained in **Figure 2** and **Table 1** seeks to balance clarity with the breadth and scope of the selected data, since there are so many disparate AM systems currently in research and production.

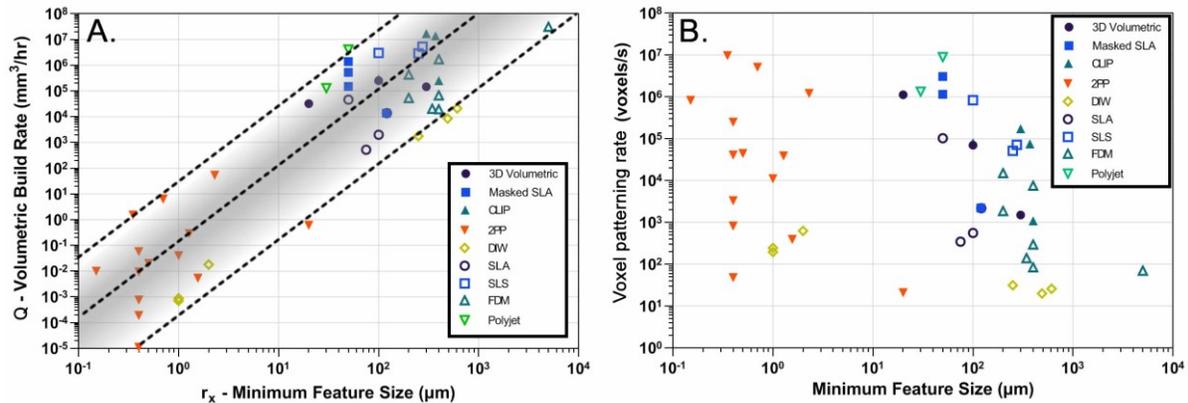

**Figure 2:** Scatter plot of selected AM technologies demonstrating of various technologies demonstrates a roughly cubic relationship between minimum feature size and volumetric build rate ($Q$). The selected datapoints were chosen to demonstrate the order-of-magnitude envelope of each AM technology; the full details for the selected AM systems can be found in **Table 1**.

To make this relationship more concrete, we consider different methods of depositing material employed by disparate classes of AM systems. FDM and DIW generally use single-point deposition via extrusion using a mechanical gantry-type motion system (such as Cartesian, Delta, Polar, etc.), and these single-point deposition systems tend to be the slowest in their class for similarly sized features. Similarly, SLS requires thermal melting/sintering of polymer, but scanned-laser systems allow faster hardware operation compared to mechanical displacement (simply due to the lower inertia/faster response of a laser gimbal vs. a physical extruder) and generally outperform FDM for a similar power rating. 2PP and Polyjet are able to achieve very fast single-voxel deposition times, as well as high parallelization factor *n*, so they achieve the greatest equivalent performance in their feature size range. SLA[22], DLP/MSLA[23,24], Polyjet and 2PP also use chemical photopolymerization, for which the effective voxel timescale is generally faster than thermal processes which rely on polymer melting/solidification. The dimensionality of this family of processes (pointwise, layer-wise, continuous) provides a scaling factor on the overall throughput. However, the increased interlayer time somewhat reduces the benefit of parallelization. The continuous versions of these processes, such as CLIP[15] and HARP[25] minimize the delays due to re-coating, but are nevertheless limited by material flow rate. Recent innovations are improving on this bottleneck.[26] Other industrial systems such as NanoScribe 2PP systems can produce objects at multiple length-scales at a roughly constant VPR, moving along the roughly diagonal (cubic) trend-line in **Figure 2A**.[27] Other recent developments in the field have enabled the possibility to print objects at one length-scale/resolution and then isotropically scale the printed objects via dehydration of printed hydrogel structures[28], or expansion of a foaming polymer[29,30].

Finally, distinct from all other forms of additive manufacturing, volumetric additive manufacturing (VAM)[31,32], of which Computed Axial Lithography[21] is an implementation, prints by illuminating all points in a volume of photosensitive resin simultaneously. The 3D distribution of absorbed energy cures an entire 3D polymer structure simultaneously, rather than layer-by-layer, within a volume of resin that is pre-assembled. Since new material is not added during the printing process, the speed of patterning is not fundamentally limited by hydrodynamic forces in the relative motion of materials. Together with its high degree of parallelization, this approach has tremendous advantages for build rate, allowing complex and soft objects to be printed very rapidly without the use of support material. Due to its simultaneous access to all spatial voxels, this printing modality is not neatly expressed by the generalized model proposed in **Eq. 3**. In VAM the projected patterns are two-dimensional, but illuminate the entire build volume, rather than a single layer. Being azimuthally-arranged, these projections thereby sample the Fourier (spatial frequency) domain of the structural information along angular 2D slices instead of the physical domain. Therefore, the print time in computed axial lithography (CAL) is governed by a special implementation of **Eq. 3**, where print time is proportional to the number of projections required and the time needed to deliver each projection. We discuss this special implementation in greater detail in the Supplementary Information and **Table S1**.

A plot to analyze a similar dataset to that shown in **Figure 2A** was constructed by Hahn et al.[17], where the authors focused attention on the voxel printing rate of an AM system compared with the voxel size and information transfer rate. The authors also maintain a continually updated version of their plot online[33]. This voxel printing rate metric we feel is a useful figure of merit for AM systems, measuring the rate of structural information processing, regardless of voxel size. These investigators defined a "fully scalable" AM technology as one which can preserve its volumetric build rate $Q$ while scaling up or down minimum feature size $r_x$. Relating this to our model, we define a *voxel patterning rate* (VPR, with units of s$^{-1}$ or Hz), for a process that has a volumetric build rate $Q$ [m$^3$/s] producing voxels with volume $v_0$ [m$^3$]. It can also be arrived at by dividing the number of simultaneous deposition paths by the time required to pattern each voxel.

$$\text{VPR} = \frac{Q}{v_0} = \frac{Q}{dr_x^3} = \frac{n}{ct_0} \qquad (4)$$

A discussion and relationship for VPR in VAM is presented in the supplemental information **Table S1**.

The plot in **Figure 2B** is generated from the same data as **Figure 2A**, deriving from **Table 1**, which includes all of the references used by Hahn et al to generate their figure. This pair of figures provides different vantage points with which to assess the performance and scalability of AM technologies, depending on which figure of merit is most valuable for a particular application.

### D.    Limitations on the Scaling of AM Technologies

The implications of the large-scale trends observed here are quite profound. The strongest of these trends is the cubic dependence of build rate on voxel size. It suggests that some of the more ambitious structures (especially those with high disparities between voxel size and build volume) may remain out of reach for conventional AM technologies. A metamaterial requiring ~10 µm features might be produced, in the most optimistic case, at ~10 cm$^3$/hr, which means many practical components requiring only 0.1 m$^3$ of such material (e.g. aerospace or automotive) would require over a year of print time. In the realm of bioprinting, if we desire to fabricate human organs or tissues with ~1 µm resolution/accuracy, we are likely to be limited to 10-100 mm$^3$/hr, therefore requiring months to years to produce a ~100-1000 cm$^3$ (kidney-to-liver scale volumes respectively).

Note that even this modest throughput is only enabled by having a large number of parallel deposition paths $n$. In processes such as DLP, CLIP, tomographic VAM and Xolography, a digital micromirror devices (DMD) is used to obtain a high degree of parallelization, as each one of the millions of active micro-mirrors can serve as an independent deposition path. Parallelization, combined with shortening the effective voxel timescale $t_0$ has proven to be a powerful strategy to increase build rate. Examples in **Table S1** show that parallelization can effectively reduce the order of minimum print time in various print modalities. However, it seems unrealistic to expect more than a 10-fold improvement beyond the current $n$ ~10$^6$ parallel deposition paths in masked/parallel SLA (driven by state-of-the-art DMD devices) or $n$ ~10$^4$ nozzles in multi-jet printing nozzles, so parallelization is approaching its limits in allowing AM technologies to scale. These limits are reflected in maximum VPR values of ~10$^7$-10$^8$ voxels/s in **Figure 2B**. Much progress can nevertheless be anticipated among AM technologies in which parallelization remains to be fully exploited, and in which the effective VPR can be enhanced, perhaps by eliminating the need to redeposit material for each layer.

Fundamentally, however, assembling a structure from voxels of uniform size via traditional AM will remain strongly constrained when the voxel size is very small, even if VPR values continue to increase. Therefore, one possible strategy to move beyond these constraints is to develop methods that take advantage of non-uniform voxel size. Such voxels would ideally be variable in size over at least 1-2 orders of magnitude, allowing more complex structural regions to be patterned in fine detail, and lower-complexity regions to be patterned with higher volumetric rate. Other potential approaches for leveraging higher VPR values from conventional AM systems are discussed in the Supplemental Information.

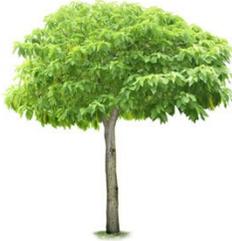

**Figure 3:** A proposed method of classifying structural manufacturing techniques with regards to the size of assemblers in relation to the finished object, and whether material is added to or removed from the structure as it progresses towards a final form.

### E. A New Framework for Manufacturing: Top-down vs. Bottom-up

As we consider what future directions may allow AM to exceed its current scaling limitations, we propose here a new framing for the total landscape of "manufacturing" processes. In order to do so, we must expand our definition of stored structural information from the digital to include the biological and genetic as well.

One may conceptualize the various paradigms for converting information into ordered matter in terms of two controlling parameters: the direction of the flow of matter (i.e. additive vs. subtractive), and the direction of the flow of information. Structural information can come from outside the assembling structure into it, which we call *"top down;"* conversely, it can be stored in some way within the structure or its constituents, which we call *"bottom up."* **Figure 3** depicts this foundational classification with illustrations. In this arrangement all existing 3D printing is found in the upper right quadrant as "top-down additive" manufacturing. Injection molding could be split into two steps, the first being the translation of digital information to a physical mold structure which is generally a "subtractive" process, and the duplication/formative step where molten polymer material is injected into a negative cavity which merely creates a copy of the pre-formed structure, but information is not being retrieved from an information processing system. Other traditional manufacturing, such as milling or machining, are clearly "top-down subtractive" processes, shown at the bottom right, represented by conventional machining. The bottom-up counterpart in additive manufacturing can be represented by the growth of living systems, represented at the upper left by a tree. Finally, the bottom-up analog of subtractive techniques in the lower left can be represented by bacterial/fungal degradation, or insect excavation of natural materials into colonies,.

With this classification scheme in mind, it may be useful to consider the "bottom-up additive" paradigm of the top left quadrant as being a promising future direction that might allow AM to exceed the scaling limitations that we have observed. In particular, a biological cell embodies the concept of a *self-replicating assembler*, which contains within itself the structural information describing the structure to be created, and uses its own mass as the build material.

An early foray down the path towards engineered self-assembly was the RepRap project[34] whose initially-stated aim was to create a self-replicating machine. The RepRap approach used a top-down 3D printer (usually FFF) to produce a kit of many of its own components, thus enabling a user to assemble a copy of the original printer using AM parts and commercially available stock materials. While the RepRap is not fully self-replicating (the printer cannot fabricate electronics, metal parts, motors, wiring and the thermal extruder), it demonstrates a device that can produce all the custom parts needed to produce a copy of itself. This paradigm captured the imagination of a sizable community of hobbyists and makers and led to a revolution in open-source AM technology which still continues to this day.

Existing research in relevant fields generally subdivides among topics such as macro-scale self-assembling structures[35,36], robotic swarms[37,38], and the mathematics of biological growth.[39] The field of AM has only begun to interface with some of these topics, particularly in the area of bioprinting, where living cells are incorporated into 3D printed structures. We anticipate that greater interconnections between these areas can bring further breakthrough advances in the scalability and functionality of AM.

The extensive and highly mature field of developmental biology has elucidated some of mechanisms behind how genetic information affects the growth, maturation, and structural formation of living systems. Contemporary research provides a starting insight into the potential scalability of bottom-up additive manufacturing into the biological domain. For instance, over the recent decades, Goriely has developed formal mathematical models for the growth of biological systems.[40] These models suggest that in some development phases of particular structures in biological systems, volumetric modes of growth are possible, and may inspire the AM engineering community in new directions.[41,42]

In related work, the possibility of self-replication of multicellular structures has recently been demonstrated by Bongard *et al.,* who showed multiple generations of progenitor-offspring cycles using frog embryonic stem cells.[43] If we look at biological growth as a "manufacturing" system, transporting new material to the self-replicating structure becomes a primary factor which limits growth rate. We will return to this subtopic at a later point.

Meanwhile, we will develop a simplified description of a "bottom-up" self-replicating assembler as a manufacturing system, using the self-replicating growth of bacterial cells as a starting point. It is important to note that while there are examples in the literature of organisms forming microfibers[44] or biofilms[45] within a top-down patterned scaffolding structure, such cells have not yet been harnessed to construct macroscopic or hierarchical multi-scale structures showing large-scale order in a bottom-up fashion. Therefore, our description is necessarily limited to best-case scaling observations and trends. Nevertheless, these considerations may complement the generalized relationships for top-down AM we have derived above, and lead to some possible directions for which to direct bio-inspired AM research.

### III. An Initial Simple Model for Bottom-Up Additive Manufacturing Systems

For a self-replicating bottom-up manufacturing system, we can largely reuse the same parameters that describe top-down systems, with some modifications. **Figure 4** shows a colony of bacterial cells, which reproduce by binary fission and themselves become part of a growing physical structure. Since each microbe acts as a self-replicating assembler, every cell can be considered as the "active voxel", and cell's cross section viewed as an independent "deposition toolpath," similar to the geometric parameters proposed in **Figure 1** above.

In the case of a self-replicating system, we can express the number of these simultaneous assemblers (i.e. simultaneous deposition paths) $n$ as a function of the current time $T$, the initial number of assemblers $n_0$, and the self-replication (doubling) time $t_d$ using a well-known relationship commonly used to describe the growth of bacterial colonies[46]:

$$n(T) = n_0 2^{\left(\frac{T}{t_d}\right)} \tag{5}$$

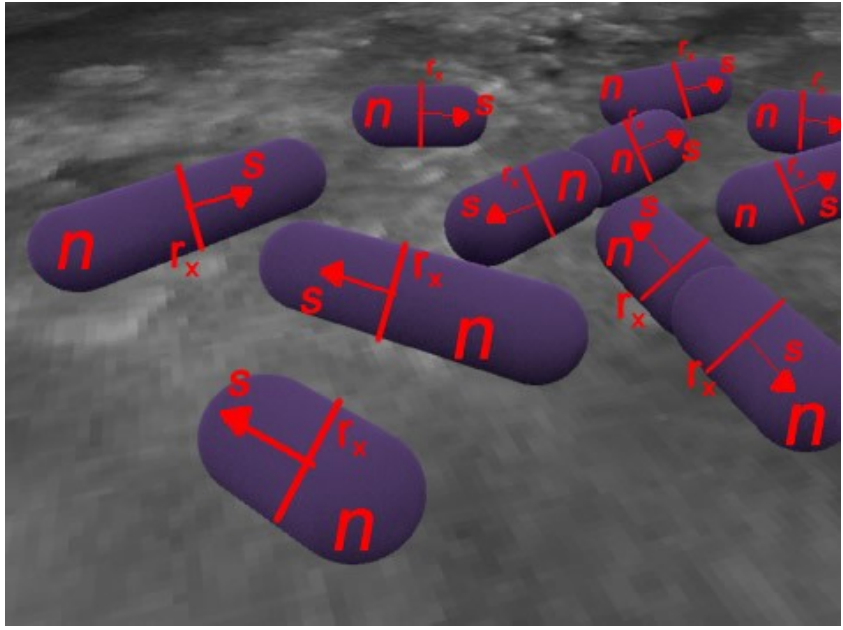

**Figure 4:** One can apply the same geometric parameters developed in the earlier section to a system of self-replicating, self assembling volumetric assemblers such as in growing *E. coli* bacteria through binary fission.

As before, the volume ($v_0$) of an individual assembler at a given time can be approximated by $v_0 = mr_x^3$, where $r_x$ is the radius or characteristic linear dimension of the bacterium, and $m$ is a geometric constant that accounts for the non-uniform growth of the individual assembler unit (rod-like vs disc-like vs spherical). Likewise, as before, we can introduce scaling coefficients such as $c$ to account for fractional packing if void space is present between assembler units (for instance, transport conduits to mimic the vasculature of living systems required for developing material). Incorporating these into the exponential growth relationship of **Eq. 5**, we obtain **Eq. 6** describing the bounding volume of the object $V_T$ which is equal to the number of assemblers as a function of time, multiplied by the individual volume of each assembler, divided by the packing fraction of the assembler units within the bounded volume:

$$V_T(T) = \frac{m}{c} r_x^3 n_0 2^{\left(\frac{T}{t_d}\right)} \tag{6}$$

As a counterpart to **Eq. 3** presented earlier, we can solve for the time $T$ required to print an object of volume $V_T$ by taking the logarithm.

$$T = t_d \log_2 \left( \frac{cV_T}{mr_x^3 n_0} \right) \tag{7}$$

From these equations, it is plain to see why a bottom-up approach may be advantageous. In bottom-up systems, the object's volume $V_T$ can increase at an exponential rate with respect to time, rather than linearly for top-down systems (as expressed in **Eq. 1 and 2**). In other words, the build time only has a logarithmic dependence on total object volume, as well as on the size of the assembler/voxel. In contrast to top-down AM, the build time is now most strongly dependent on the self-replication time $t_d$. In the case of bacteria, doubling times can be as short as 20-30 minutes under ideal laboratory conditions, but may be tens of hours under more typical "wild" conditions.[47] For mammalian cells (including the frog embryonic stem cell spheroids used by Bongard and colleagues) doubling time is similarly on the order of 10s of hours.[43]

Given our simple model, for a notional fungal or bacterial colony with initial parameters of $r_x$=5 µm, $t_d$=30 min, and $n_0$=100 cells, the logarithmic growth rate implies that a structure volume of 1 m³ could be constructed in approximately 24 hrs of exponential growth. Of course, no microorganism colony is able to sustain exponential growth for more than a few hours without active management of nutrient supply and waste removal.[48–50] Without differentiation into multiple cell types and tissues, and without hierarchical organization, solid biological structures (e.g. solid tumors) that grow by simple exponential replication are limited to the length scale of diffusive transport along at least one spatial dimension, which is on the order of ~100 µm.[40,51] Cell differentiation, structurally-directed growth, the development of hierarchy and organization, and the deposition of non-living structural materials are what causes growth rates in more complex living systems to be far slower than the exponential growth limit.

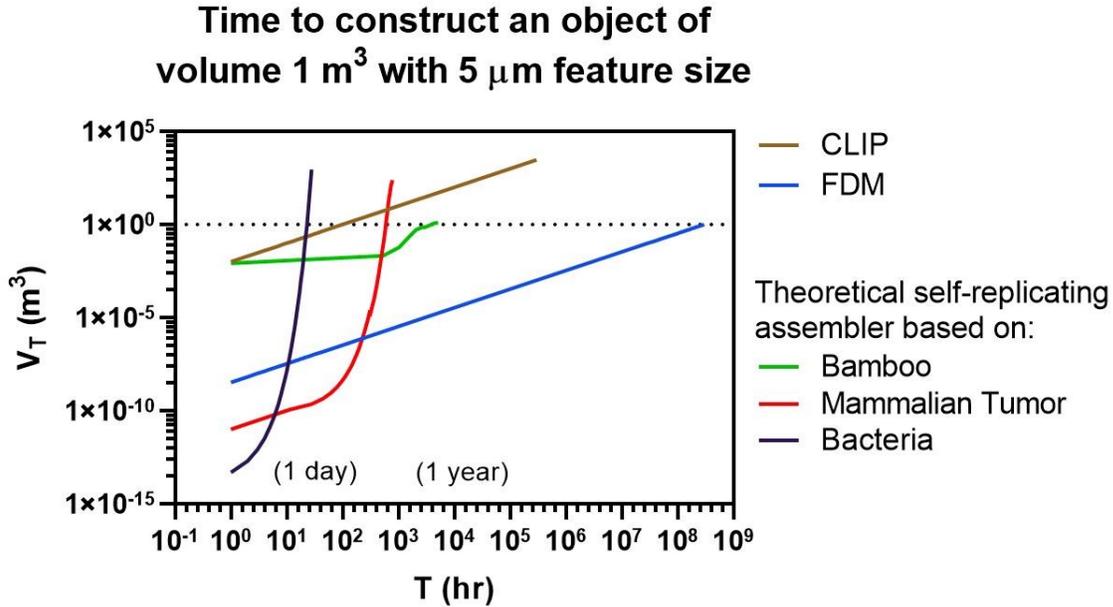

**Figure 5:** Volume of self-replicating vs top-down assembled systems with respect to time. Using the model in **Eq. 6 and 7** we obtain a volume of 1 m³ at approximately 10 hrs with a bacteria-based self replicating system with parameters $r_x$=5 µm. Other theoretical self-replicating assemblers based on different types of engineered biological systems are presented for perspective of real world growth dynamics.

Nonetheless, using **Eq. 7** and data prepared from contemporary literature, we can contrast build times for theoretical self-replicating/self-assembling systems with conventional AM systems of comparable resolution. It is important to highlight that even though the bacteria, bamboo, and mammalian tumors all began with a starting population of $n_0 = 100$ and were prepared from different data sources (our **Eq.7** theoretical approximation vs literature), the graphs show marked similarity in shape. It is also significant that to note the bacteria's doubling time ($t_d = 30$ min), the mammalian tumor's doubling time ($t_d = 17$ hr), and the bamboo's doubling time ($t_d \approx 300$ hr) produce significantly different build times for a 1 m³ structure despite possessing the same initial population $n_0 = 100$, which is anticipated by our generalized model. Raw data as well as details on the reduction and preparation of data on the hypothetical biological systems for **Figure 5** are elaborated in **Table S2-3.**

We can compare these results with our model's implications for the limitations of conventional AM systems using **Eq. 3**: a hypothetical CLIP system with a 1 m² display surface and a display with $(2\times10^5) \times (2\times10^5)$ pixels (5 μm pixel width) printing at 10 mm/hr would require 100 hours to print such a structure, assuming the display could be cooled adequately. A similarly capable FDM system operating with a single independent print head of 5 μm diameter at the highest recorded FDM print speed would require 34,000 years of operation to print such a structure. While all of these calculations put aside practical nuances for the sake of simplicity, even with an oversimplified model of idealized systems, the advantages of exponentially replicating systems are difficult to ignore.

At this time it may be helpful to conceptualize the framework for a hybrid approach merging top-down and bottom-up AM, in order to highlight the advantages of both and overcome their limitations. A top-down assembler could be used to provide hierarchical organization and vascular structure to enable nutrient transport and waste removal. The voxel size $r_x$ of this top-down system may be sufficiently large (10-1000 μm) to enable relatively fast construction. Meanwhile, the feedstock material that this assembler deposits might contain self-replicating assemblers (e.g. microbial or fungal cells) to generate bottom-up growth on a microscopic length-scale, perhaps programmed for μm-scale precision in their self-assembly, defining the ultimate material (or other) properties of the overall structure or its regions. In the realm of bioprinting, researchers have begun incorporating living microbial, fungal, and mammalian cells into AM feedstocks[52,53], though programming their assembly into finer-scale structures is just crossing the threshold of feasibility.[54,55] This approach will require continuing significant advances in synthetic biology for engineered microorganisms[56,57] as well as continued progress in developing living feedstock materials for bioprinting. Interim approaches such as the use of phototaxis[58] or magnetotaxis may be employed to pattern and direct the replication of cells in a similar fashion.

The bottom-up technology we envision does not necessarily have to be purely biological; this could be an engineered system that mimics certain attributes of biological growth but in a more directly programmable way. Such advances may likewise require hybrid approaches, blending "top down" miniaturization of electronic and mechanical robotic elements[59] with "bottom-up" engineered molecular structures.[60] The biggest limitation to incorporating bottom-up additive manufacturing paradigm is the ability to direct a bottom-up system to recursively assemble a final desired structure. Machine learning and artificial intelligence have recently proved key in predicting once "unsolvable" folded protein structures from amino acid sequences[61], and may well hold the key to achieving such similar predictive feats of bottom-up directed cellular assembly.

## IV.    Outlook for biological self-assemblers as potential manufacturing systems

In order to move beyond the cubic scaling-law limitations of top-down AM, a bottom-up self-replicating paradigm may become a productive direction for innovation. At a minimum, the basic theoretical model (**Eq. 7**) sketched here can inspire and motivate this direction of investigation. Today, this model cannot be refined to take into account the complex growth conditions, nor the vast array of parameters and configurations which self-replicating bottom-up assemblers might adopt. We do not yet possess the means to harness the self-replication tools which evolution has taken billions of years to develop. Likewise, the computational tools necessary to develop self-assembling, self-replicating assembled features from living systems are still forthcoming.

However, by exploiting the massive parallelization that can be accessed through bottom-up assembly, one could envision that such systems may eventually come to complement or even supplant existing manufacturing methods. In the same way that subtractive mechanical fabrication, mass production and automation eventually replaced artisan manufacturing using hand tools, self-directed assembly may hold the potential to supplant top-down mechanical fabrication. Assembly of a biopolymer structure in the desired geometry might be as simple as planting a seed and allowing a self-assembling system to absorb nutrients from the soil and energy from the sun.

While the fastest growing materials may be biopolymers and other high-water content bacterial structures, more robust materials such as wood, ceramic, composites and even metals may be biosynthesized and structurally patterned via bottom-up manufacturing as well, perhaps with a much slower growth rate (i.e. voxel patterning rate). Such a paradigm might allow for massively parallel "farming" of objects, independent of centralized production and tooling. While each individual structure may take years to develop, the advantage of unlimited parallelization without human-dependent serial labor would vastly reduce the per-unit build time, only requiring the addition of land and feedstock to construct additional units. Such a paradigm would be ideal for in-situ utilization of resources in austere environments, or construction of large structures without human intervention, or transportation of large tooling to such locations (i.e. planetary colonization). While top-down manufacturing will likely remain the preferred framework for many industrial processes, to build next generation, multi-scale structures with high resolution in a practical timescale, the bottom-up approach may enable the critical breakthroughs that take manufacturing beyond the current frontier.

| Technology class | System description | Min feature size (µm) | Volumetric build rate (mm$^3$/hr) | Notes | Ref |
|---|---|---|---|---|---|
| 3D Volumetric | LLNL/MIT prototype* | 100 | 252000 | Max volume / min exposure of current config 0.7 cm$^3$/sec | 21 |
| | UC Berkeley Prototype | 300 | 146149 | | 31 |
| | UC Berkeley micro-CAL | 20 | 32311 | trifurcated channel, silica, 20 um resolution, 30s print time, 3.5mm diam x 7mm height | 62 |
| | Xolographic | 21 | 198000 | 21 µm pixels, 55 mm$^3$/s max absorbtion | 63 |
| CLIP | iCLIP | 30 | 294912 | 80 mm/hr with 76.8x48 mm array | 26 |
| | Carbon3D | 400 | 250000 | Shoe 100x25x100 mm/hr | 15 |
| | | 100 | 288000 | Argyle 24x25x500 mm/hr | 15 |
| | | 50 | 14400 | Paddles 24x24x25 mm/hr | 15 |
| | HARP | 300 | 16700000 | High area rapid printing | 25 |
| 2PP DLW | Nanoscribe | 0.15 | 0.01 | Using 63x objective | 21 |
| | | 0.5 | 0.02 | Using 25x objective | 21 |
| | Parallelized 2PP | 0.7 | 6.31 | | 17 |
| | | 0.35 | 1.51 | | 17 |
| | 2PP NIL | 0.4 | 0.00076 | Fabrication of Nano-Imprint Lithography | 64 |
| Polyjet | Stratasys Objet 5000 Connex 1 | 30 | 129117 | Vendor information release, 2014 | † |
| | HP JetFusion 3D 4210 | 50 | 4016845 | Vendor information release, 2017 | † |
| DIW | J. Lewis (UIUC) | 1 | 0.000707 | 1um filaments at 250 µm/s | 12 |
| | J. Lewis (UIUC) | 2 | 0.0181 | 1um filaments at 250 µm/s | 13 |
| | LLNL DIW | 610 | 21041 | 20 mm/s | 65 |
| | LLNL DIW | 250 | 1767 | 10 mm/s | 21 |
| | MEMS DIW | 1 | 0.00088 | DIW assembly of Si-MEMS photonic crystals | 12 |
| SLA | Autodesk Ember | 50 | 46080 | 50 µm XY, 50 µm layer height, 64x40 mm area, 18 mm/hr | † |
| | FormLabs Form2 | 100 | 2000 | 2 cm/hr, 10x10 mm area, rook example part | † |
| | 3D Systems Projet 7000 HD | 75 | 522 | Vendor information release, 2012 | † |
| SLS | 3DS sPro 230 HD-HS | 100 | 3000000 | 3.0 L/hr, 100 µm layer height and 100 µm XY res | † |
| | EOS P770 | 250 | 2880000 | 400 µm laser spot x 100 µm layer height x 20 m/s scan speed | † |
| | EOS P110 | 274 | 5260000 | With PA 2200 "Top Quality" | † |
| FDM | BAAM | 5000 | 13600000 | 0.6m$^3$ in 44h | 66 |
| | Prusa | 200 | 54000 | 0.4mm nozzle, 15mm3/s (standard prusa) | † |
| | Voron | 200 | 432000 | 0.4mm nozzle, 120mm3/s high speed voron | † |
| | Voron | 400 | 1728000 | 0.8mm nozzle, 480mm3/s high speed voron | † |
| | Univ. Maine 3DP | 5000 | 31500000 | ~2.268m3/72hr | † |
| | Stratasys Mojo | 343 | 20425 | | † |
| | Stratus Fortus 360MC | 400 | 68129 | | † |
| | Ultimaker 2 | 400 | 19430 | | † |
| MSLA | LED-based Projection SLA | 120 | 13770 | Early research prototype | 24 |
| | LAPµSL | 50 | 270 | | 23 |
| | LAPµSL | 100 | 960 | | 23 |
| | LAPµSL | 120 | 1093 | | 23 |
| | LAPµSL II | 200 | 13670 | | 23 |
| | LAPµSL II | 120 | 13853 | | 23 |
| | Anycubic Mono | 50 | 520000 | | † |
| | Anycubic Mono X | 50 | 1382400 | | † |

**Table 1**: Selected references and systems highlighted in Figure 1 and associated references.
†Specifications obtained from relevant company website.
*Limited geometry due to three intersecting beams